\documentclass[twoside,11pt,francais,notilepage]{article}

\usepackage[francais]{babel}
\usepackage[applemac]{inputenc}
\usepackage{authblk}
\usepackage{times}

\usepackage{graphicx}
\usepackage{amsmath}
\usepackage{amssymb}

\usepackage{latexsym}

\usepackage{color}

\definecolor{gris}{gray}{0.7}
\definecolor{gris2}{gray}{0.5}
\definecolor{gris3}{gray}{0.35}
\definecolor{vert}{rgb}{0.,0.8,0.5}
\definecolor{bleuclair}{rgb}{0.,0.6,0.8}
\definecolor{orange}{rgb}{1.,0.5,0.}
\definecolor{violet}{rgb}{0.4,0.,1.}

\newcommand{\HI}{$\mathrm{H_I}$ } 

\begin{document}
\begin{center}
%% Titre 
{\Large \bf BAORadio: Cartographie 3D de la distribution de gaz H$_\mathrm{I}$ \\ dans l’univers} \\[5mm]
%% Liste d'auteurs 
R. Ansari$^1$,   
J.E. Campagne$^1$, 
P. Colom$^2$,
C. Magneville$^3$,
J.M. Martin$^2$, \\
M. Moniez$^1$
J. Rich$^3$
C. Yèche$^3$  \\[4mm]
\parbox{0.95\textwidth}{\small 
1- LAL, Universit\'e Paris-Sud,  CNRS/IN2P3, UMR 8607, F-91898 Orsay Cedex, France \\
% 1- LAL, Univ Paris-Sud, CNRS/IN2P3, Orsay, France \\  % pour le LAL 
2- GEPI, UMR 8111, Observatoire de Paris, 61 Ave de l'Observatoire, 75014 Paris, France \\
3- CEA, DSM/IRFU, Centre d'Etudes de Saclay, F-91191 Gif-sur-Yvette, France \\[3mm]
Mots-clefs: cosmologie, énergie noire, cartographie à 21 cm, radio-interférométrie 
} \\[3mm]
%%% Resume
\parbox{0.8\textwidth}{
{\bf Résumé:} La cartographie de l’univers en radio, à travers l’observation de la  raie à 21 cm de l’hydrogène atomique, constitue une approche complémentaire aux relevés optiques pour l’étude des grandes  structures, et en particulier des oscillations acoustiques baryoniques (BAO).
Nous proposons une méthode originale de la mesure de la distribution de l'hydrogène atomique 
neutre à travers une cartographie à 3 dimensions de l'émission du gaz à 21 cm, sans passer 
par l'identification des sources compactes (galaxies \ldots). Cette méthode nécessite un 
instrument à grande sensibilité et ayant une grande largeur de bande instantanée ($\gtrsim$ 100 MHz), alors qu'elle peut se contenter d'une résolution angulaire moyenne (10 arcmin). 
L'instrument devra avoir une surface de collection de quelques milliers de m$^2$ et quelques centaines de lobes simultanés. Ces contraintes peuvent être satisfaites avec 
un réseau dense de récepteurs en mode interférométrique ou un réseau phasé au foyer 
d'une grande antenne.
}
\end{center}
%%%%
%%%%%%%
\section{Introduction}
Les questions concernant la nature et les propriétés de l’énergie noire se trouvent au cœur de la cosmologie et de la physique aujourd’hui. L’énergie noire est une composante mystérieuse responsable de l’accélération de l’expansion de l’Univers. Les oscillations acoustiques du plasma photons-baryons antérieur au découplage sont à l'origine des modulations du spectre de puissance de la distribution de matière, appelées BAO. Ces oscillations acoustiques ont clairement été mises en évidence dans les anisotropies du fond diffus \cite{wmap.11}. La mesure de l’échelle de longueur des oscillations à différents redshifts est une sonde cosmologique de type {\it règle standard}, considérée comme l’une des méthodes les plus robustes pour  contraindre l’équation d’état de l’énergie noire \cite{bao.07}.

La distribution de matière dans l’univers peut bien sûr être observée en optique, en utilisant les galaxies comme traceur. L’observation radio de l’émission à 21 cm de l’hydrogène neutre est une méthode complémentaire pour déterminer la distribution de matière. U-L Pen et J. B. Peterson ont proposé un projet radio en vue de cartographier la distribution de l’hydrogène atomique à travers son émission à 21 cm (transition hyperfine, 1.42 GHz à z=0). Il est important de noter que la méthode d’observation envisagée ici repose sur la cartographie de l’émission totale à 21 cm en fonction du décalage vers le rouge, sans rechercher l’observation des sources individuelles (galaxies…) \cite{peterson.06} \cite{ansari.08} \cite{chang.08}. 
L’objectif est de produire une cartographie à 3 dimensions de l’émission HI (2 angles + fréquence (~redshift)), avec une résolution angulaire d’une dizaine de minutes d’arc et de quelques dizaines de kHz en fréquence ($dz/z <10^{-3}$).

Un programme de recherche et développement en électronique pour l’interférométrie radio a été initié par le LAL (IN2P3) et 
l’Irfu (CEA) fin 2006, auquel participent également l’observatoire de Paris (GEPI et station de Nançay) depuis l’été 2007. 
Une chaîne complète comprenant des modules analogiques de conditionnement des signaux radio, des modules de 
numérisation et de filtrage numérique et un système d’acquisition et de traitement haute performance, a été développée 
grâce au soutien de P2I, de l’Irfu (CEA), de l’IN2P3 (CNRS) et du PNCG. 
Cette chaîne est utilisée pour caractériser les prototypes de réflecteurs cylindriques 
développés par l'université Carnegie-Mellon (CMU) à Pittsburgh.
Le projet FAN (Focal Array at Nançay) utilise également la chaîne électronique et informatique développée pour BAORadio. 
Ce projet a pour but le développement et la caractérisation d’un prototype de réseau de récepteurs au foyer du NRT (réseau phasé) 
et pourrait déboucher à terme sur un système multi lobes qui augmenterait de plus d’un ordre de grandeur la sensibilité du 
radio télescope de Nançay en mode cartographie \cite{martin.11}.

Les enjeux scientifiques et des difficultés spécifiques de la méthode d’observation envisagée, la cartographie 3D à travers 
la mesure de la température de brillance en fonction de la fréquence {\it (Total Intensity Mapping)}
 seront présentés dans le paragraphe \ref{sectim}. 
Nous discuterons en particulier le  niveau de sensibilité nécessaire pour la mesure du spectre des inhomogénéités de densité P(k) et 
les difficultés de séparation du signal cosmologique des avant-plans radio. Les développements techniques qui ont été menés depuis 
2007 seront présentés dans le paragraphe \ref{secdevelec}. Les résultats obtenus  lors des tests de qualification 
de l'électronique à Nançay (NRT) et sur les prototypes de réflecteurs cylindriques de Pittsburgh seront brièvement discutés.

\section{Cartographie 3D de l'émission à 21 cm}
\label{sectim}
L'étude des propriétés statistiques de la distribution de matière dans l'univers permet de tester la validité 
du modèle cosmologique et peut être utilisé pour déterminer les valeurs des paramètres cosmologiques, 
en particulier celles des densités cosmiques. On détermine la distribution cosmique de matière, dominée
par la matière noire en utilisant des traceurs, les galaxies dans la majorité des relevés. 
Ces galaxies sont en général détectées à travers leur émission optique, la position étant alors
déterminée par imagerie optique, et le décalage vers le rouge ($z$) par spectroscopie. 
Une méthode similaire peut être utilisée dans le cas de l'observation des galaxies et des nuages 
compact de gaz en radio, à travers l'émission à 21 cm. Le décalage vers le rouge de la source 
est alors déterminé directement en comparant la fréquence de l'onde radio observée $\nu$ à celle 
de la raie à 21 cm ($ z = \frac{1420 \mathrm{MHz}}{\nu} - 1$). Les applications cosmologiques de 
l'observation des galaxies à 21 cm sont par exemple discutées dans la référence \cite{ska.science}.

Mais la faible luminosité des galaxies à 21 cm limite la possibilité de détection de ces objets 
au voisinage de notre Galaxie avec les instruments actuels. 
Une antenne d'une surface de $A \sim 10^4 \mathrm{m}^2$ comme celle du grand radiotélescope de
Nançay (NRT) ne permet la détection des galaxies à 21 cm ($M_{H_I} \sim 10^{10} M_\odot$) 
que jusqu'à $z \lesssim 0.1$ avec un temps d'intégration d'une heure ($S_{lim} \sim 1 mJy$). 
La détection des galaxies en radio à 21 cm à des distances cosmologiques ($z \gtrsim 1$) 
ne peut donc être d'envisagée qu'avec des surfaces de l'ordre de $A \sim 10^6 \mathrm{m}^2$. 

Le spectre des inhomogénéités de densité $P(k)$ est l'un des outils mathématiques les plus utilisés 
pour quantifier et analyser les propriétés de la distribution de matière. Si on note $\rho(\vec{r})$ la densité 
de matière dans le cosmos et $F(\vec{k})$ la transformé de Fourier du champ des  inhomogénéités de densité $\delta(\vec{r})$:
$$ \delta(\vec{r}) = \frac{\rho(\vec{r})}{\bar{\rho}} -1 \hspace{2mm}  \rightarrow TF \rightarrow \hspace{2mm} F(\vec{k})$$ 
Si on admet l'isotropie à grande échelle de la distribution de matière, les propriétés statistiques 
de la distribution de matière peuvent être résumées dans le spectre de puissance $P(k)$ 
$$ P(k) = < | F(\vec{k}) |^2 > \hspace{3mm} \mathrm{avec} \hspace{3mm} k = |\vec{k}| $$

\begin{figure}
\centering
\mbox{
\vspace*{-10mm}
\includegraphics[width=0.9\textwidth]{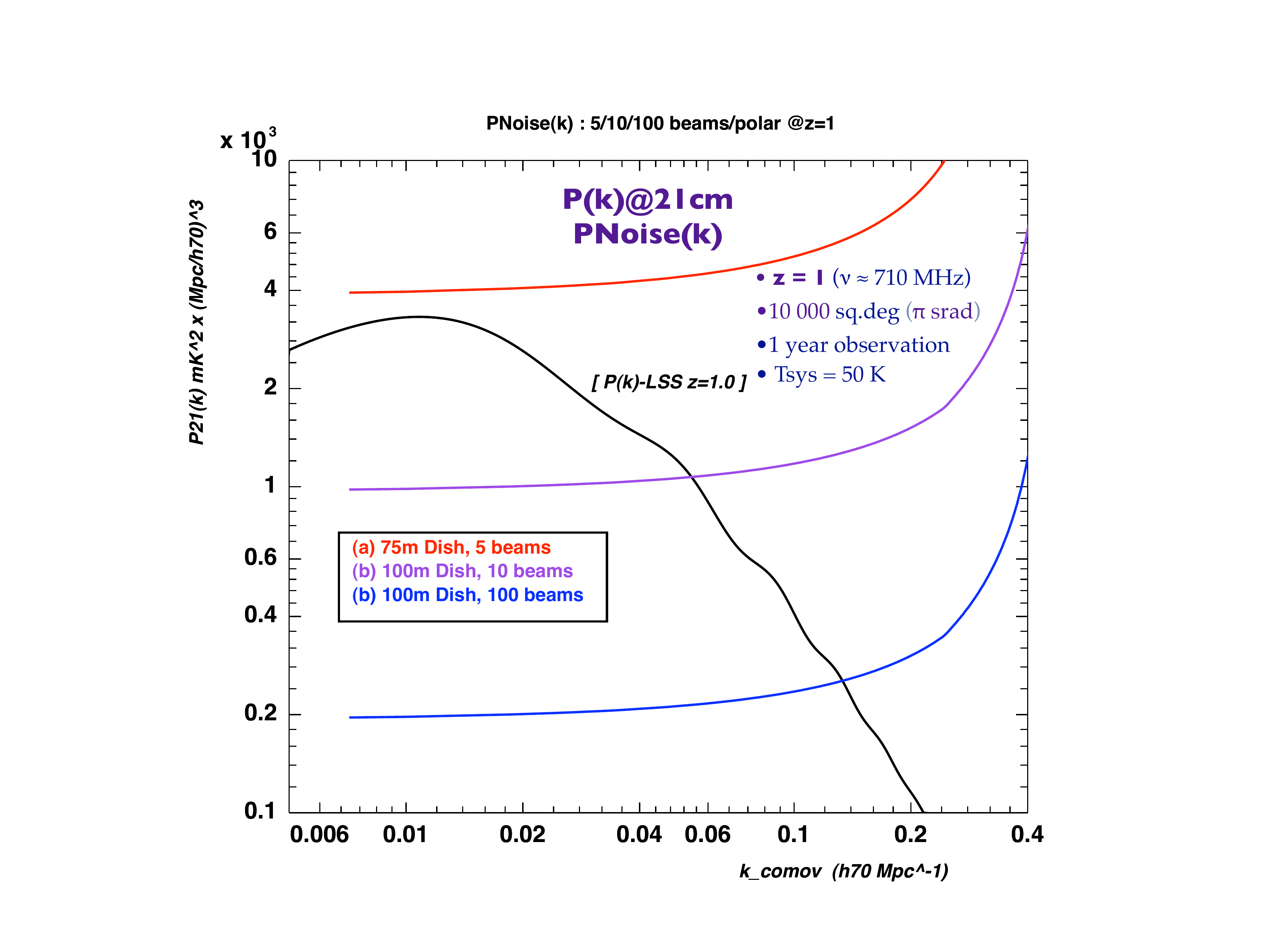}
}
\vspace*{-10mm}
\caption{ Sensibilité d'un instrument radio pour la mesure du spectre des fluctuations de température 
de brillance de l'émission \HI à $z=1$. Observation du quart du ciel ($\pi$ stéradians) pendant une année. 
Trois configurations instrumentales, avec un niveau de bruit $T_{sys} = 50 \, \mathrm{K}$: 
(a) antenne de D=75 mètres de diamètre, équipée d'un instrument focal (FPA) avec 5 lobes, 
(b) D=100 mètres et 10 lobes, (c) D=100 mètres et 100 lobes \cite{ansari.11}. }
\label{figsenspk}
\end{figure}

Les Oscillations Acoustiques Baryoniques (BAO) correspondent aux empreintes laissées dans la distribution 
de matière par les ondes de pression (son) qui se propageaient dans le plasma photon-baryon 
avant le découplage. Les BAO se manifestent comme une légère modulation ($\lesssim 10 \%$) 
du spectre de puissance $P(k)$ pour $k_n \sim \frac{2 \pi \, n}{150 \mathrm{Mpc}}$. Cette échelle 
de longueur (150 Mpc) correspond des structures dans le plan transverse ayant une taille angulaire 
de $\sim 1$ degré. 
La température moyenne de brillance à 21 cm et le niveau des fluctuations attendues sont 
inférieurs au milli-Kelvin.  
Un instrument radio de taille modeste ($D \lesssim 100 \mathrm{m}$) 
permettrait donc de résoudre ces structures. Par contre, il faut un instrument ayant une grande 
sensibilité à la température de brillance pour éviter que le signal cosmologique soit noyé 
par les fluctuations du bruit instrumental, quantifié par la température système ($T_{sys}$). 
La détermination du spectre de puissance nécessite en effet le sondage d'un grand volume
de l'univers afin de limiter la dispersion due au nombre de modes mesuré 
{\it (Sample variance) }. Il faut donc réaliser un relevé couvrant une fraction 
notable du ciel ($\sim 10000 \, \mathrm{deg}^2$) et une tranche en décalage vers le rouge 
$\Delta z \gtrsim 0.2$. Ces contraintes peuvent être satisfaites par un instrument 
ayant un grand champ de vue instantané (10-100 deg$^2$) et à large bande ($\gtrsim 100-200 \mathrm{MHz}$). 
La cartographie à 3 dimensions (2 angles, fréquence/redshift) de la température de brillance 
avec une résolution angulaire $\delta \Omega \sim \mathrm{10 arcmin \times 10 arcmin}$, et 
une résolution en fréquence $\delta \nu \sim \mathrm{100 kHz}$ pourrait être utilisée 
pour la détermination du spectre de puissance de la distribution de l'hydrogène atomique. 
La figure \ref{figsenspk} montre la contribution du bruit instrumental, pour plusieurs 
configurations instrumentales, avec 5,10 et 100 lobes simultanés, comparée au spectre 
de puissance attendu de la température de brillance de l'émission à 21 cm à $z=1$. 

\begin{figure}
\centering
\mbox{
\vspace*{-10mm}
\includegraphics[width=\textwidth]{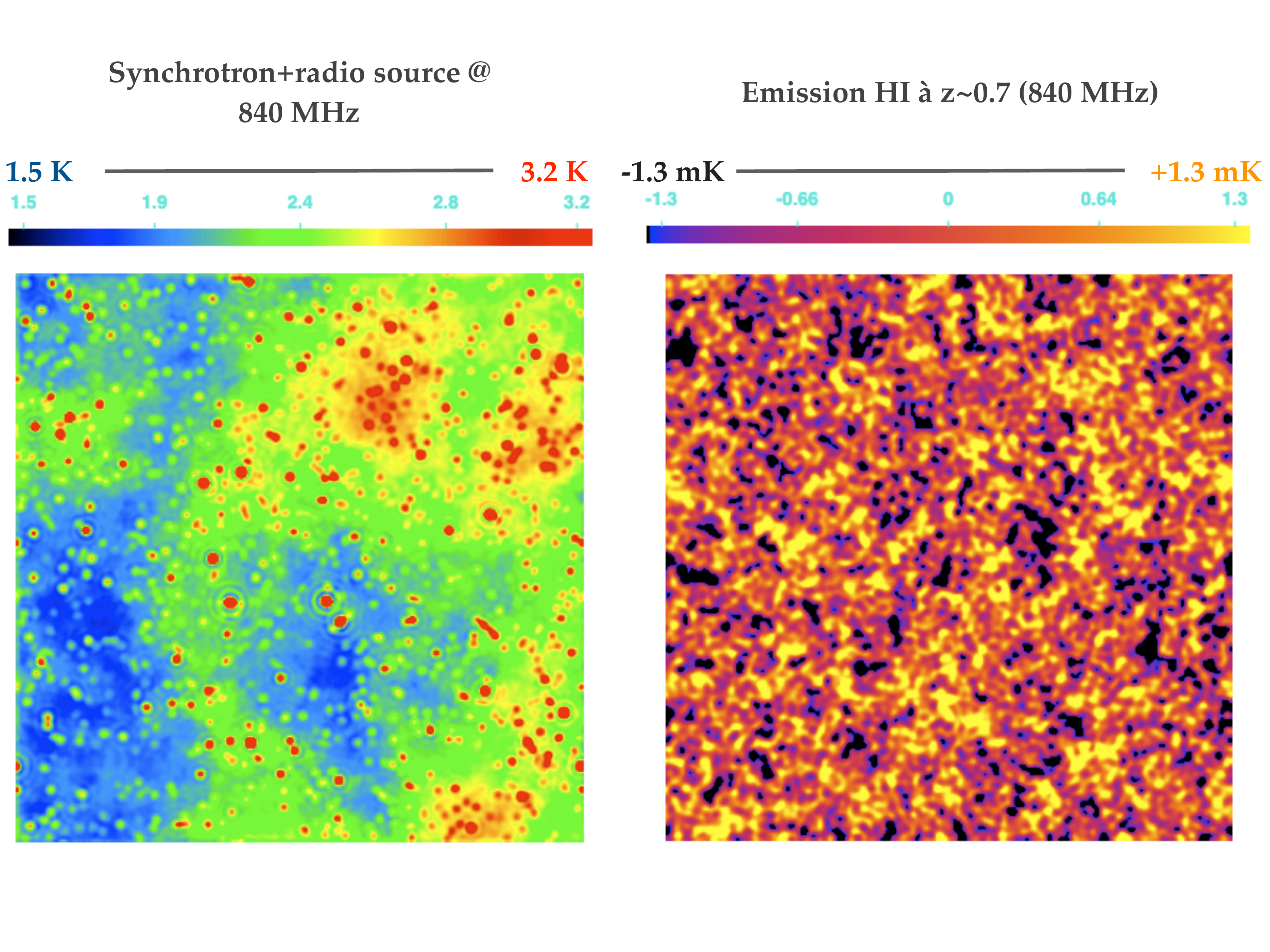}
}
\vspace*{-20mm}
\caption{Gauche: température de brillance de l'émission synchrotron et des sources radio continuum à 840 MHz 
(échelle des couleurs en Kelvin), 
pour une zone de $30 \times 30$ deg$^2$, parmi les plus froides de la Galaxie.  
Droite: température de brillance attendue pour l'émission à 21 cm de l'hydrogène atomique à z=0.7 
(échelle des couleurs en milli-kelvin)  \cite{ansari.11} }
\label{figradsrc}
\end{figure}

Le niveau du bruit instrumental ne représente pas le seul défi expérimental de la cartographie 3D de
l'émission à 21 cm. En effet, les avant-plans radio dus à l'émission synchrotron de la Voie Lactée 
et des radio sources sont plus d'un millier de fois plus intenses que l'émission de l'hydrogène 
atomique neutre. La figure \ref{figradsrc} montre le niveau des variations de température de brillance 
attendu pour la raie à 21 cm $(\nu \simeq 840 \mathrm{MHz}, z \sim 0.7)$ comparées 
à celles dues au synchrotron galactique et aux sources radio. 
Néanmoins, des études préliminaires montrent qu'il devrait être possible de séparer 
les deux contributions en utilisant la forme du spectre en fréquence  des émetteurs radio \cite{ansari.11}.
En effet, le synchrotron galactique et les radio sources ont des spectres d'émission en loi de puissance $\propto \nu^\beta$.  
Cette caractéristique peut être mise à profit pour isoler le signal \HI cosmologique.

\section{L'électronique BAORadio}
\label{secdevelec}

L'originalité du système électronique développé réside dans sa conception  
entièrement numérique, et dans l’utilisation de circuits câblés à base de FPGA afin d’obtenir une puissance de traitement importante, avec un coût et une consommation modérés. La gamme de fréquence 0.5-1.5 GHz est découpée en plusieurs bandes de ~250 MHz de large, et le signal analogique provenant des récepteurs est numérisé à 500 MHz, après une étape d'amplification et de filtrage. Ce flot numérique est ensuite traité par les différents étages de la chaîne électronique et d’acquisition. 
D'autres développements de systèmes numériques pour des instruments radio \cite{dsp.radio}  
ont été réalisés ou sont  en cours: CASPER/ROACH à Berkeley \cite{casper}   et Uniboard en Europe 
\cite{uniboard}. 

Des éléments de la chaîne électronique et le système d’acquisition ont été conçus et réalisés durant les trois dernières années 
(2007 à 2009). Outre des tests de mise au point en laboratoire, le programme de travail comprend des tests de qualification effectués en situation réelle auprès du radiotélescope de Nançay et du prototype de Pittsburgh. 
\begin{enumerate}
\item Module analogique d’amplification, de filtrage et de décalage en fréquence. Ces modules, conçus et réalisés par l’IRFU, permettent le découpage du signal en bande de 250 MHz.
\item  Système de distribution d’horloge et de signaux de synchronisation (DCLK). Celui-ci, également sous la responsabilité de l’IRFU, permet d’assurer le synchronisme temporel avec la précision nécessaire.
\item Une carte de numérisation à 4 voies, capable de numériser les signaux à 500 MHz a été conçue et réalisée par le LAL. Cette carte est équipée de circuits programmables puissants, capable d’effectuer au vol le filtrage numérique des signaux pour les séparer en composantes quasi monochromatiques (FFT). Les données brutes ou après filtrage (FFT) sont transmises par des liens optiques haut débit (5 Gbits/s) vers les ordinateurs d’acquisition, ou vers un système dédié de corrélateur et de reconstruction de lobe. Le code embarqué (firmware) de filtrage numérique est réalisé par l’IRFU et le LAL.
\item Un module de réception des données, à la norme PCI-Express, capable de soutenir des débits de plusieurs centaines de méga-octets par seconde permet le transfert des données vers la mémoire des ordinateurs d’acquisition. Le code embarqué de ce module est également développé au LAL. 
\item Un système d’acquisition et de traitement au vol des données sur des grappes de PC a été développé sous la responsabilité du LAL, en collaboration avec l’IRFU. 
\item La conception et la réalisation d’un prototype d’un système de corrélateur/synthétiseur de lobes d’antenne à base de FPGA est en cours à l’IRFU. Les premiers tests de ce prototype  sur le ciel devront avoir lieu dans les prochains mois.
\end{enumerate}
Les premiers tests de la chaîne électronique ont été effectués avec succès à Nançay au foyer du radiotélescope décimétrique (NRT) en juillet 2008. D’autres campagnes de tests ont été menées durant l’hiver et le printemps 2009. Ils ont permis en particulier de tester des versions améliorées des codes embarqués (firmware FFT) et de l’acquisition multi-voies. Plusieurs campagnes de  mesure en mode interférométrique ont été effectuées sur le prototype des réflecteurs cylindriques de Pittsburgh en Juin 2009, novembre 2009 
et décembre 2010. Un programme d'observation des amas de galaxies utilisant les récepteurs 
cryogénique du NRT et l'électronique BAORadio, ainsi que les mesures avec le prototype FAN 
ont débuté fin 2010 à Nançay.

\begin{figure}
\centering
\vspace*{-15mm}
\mbox{
\includegraphics[width=0.85\textwidth]{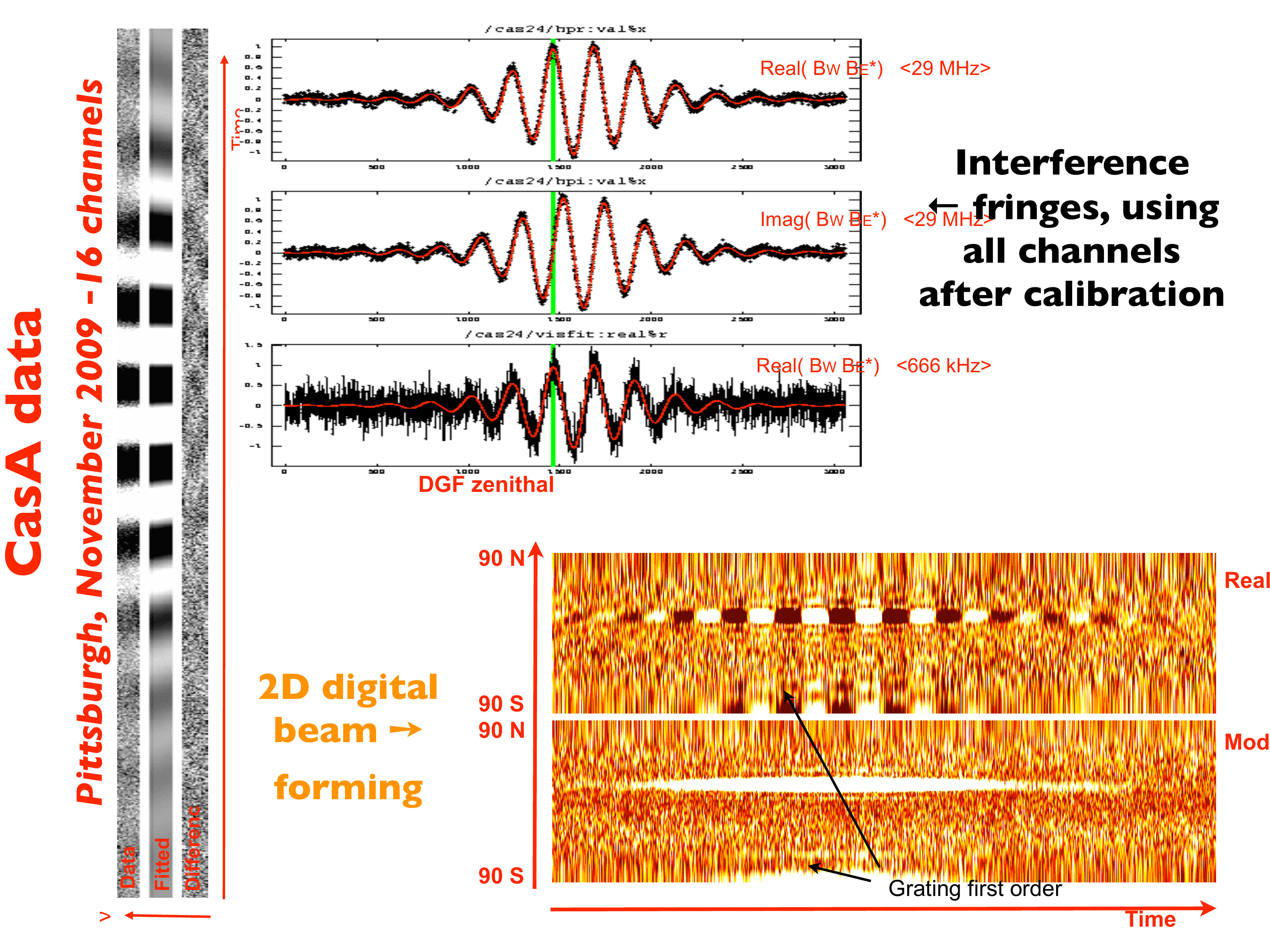}
}
\vspace*{-7mm}
\caption{Signaux d’interférence et synthèse de lobe, données Pittsburgh de Novembre 2009}
\label{figpittsnov09}
\end{figure}

Les observations en mode interférométrique réalisées
sur le prototype de PittsBurgh en Novembre 2009 ont permis
de vérifier le fonctionnement de la cha\^{\i}ne électronique:
l'observation de sources brillantes (CasA, CygA, Soleil)
a permis la calibration des voies d'électronique,
la reconstitution de lobes numériques (de type réseau phasé)
ainsi que la vérification de la stabilité temporelle relative
des voies d'électronique. Une partie des résultats obtenus est résumée sur la figure \ref{figpittsnov09}.

\begin{figure}
\centering
\mbox{
\vspace*{-15mm}
\includegraphics[width=0.8\textwidth]{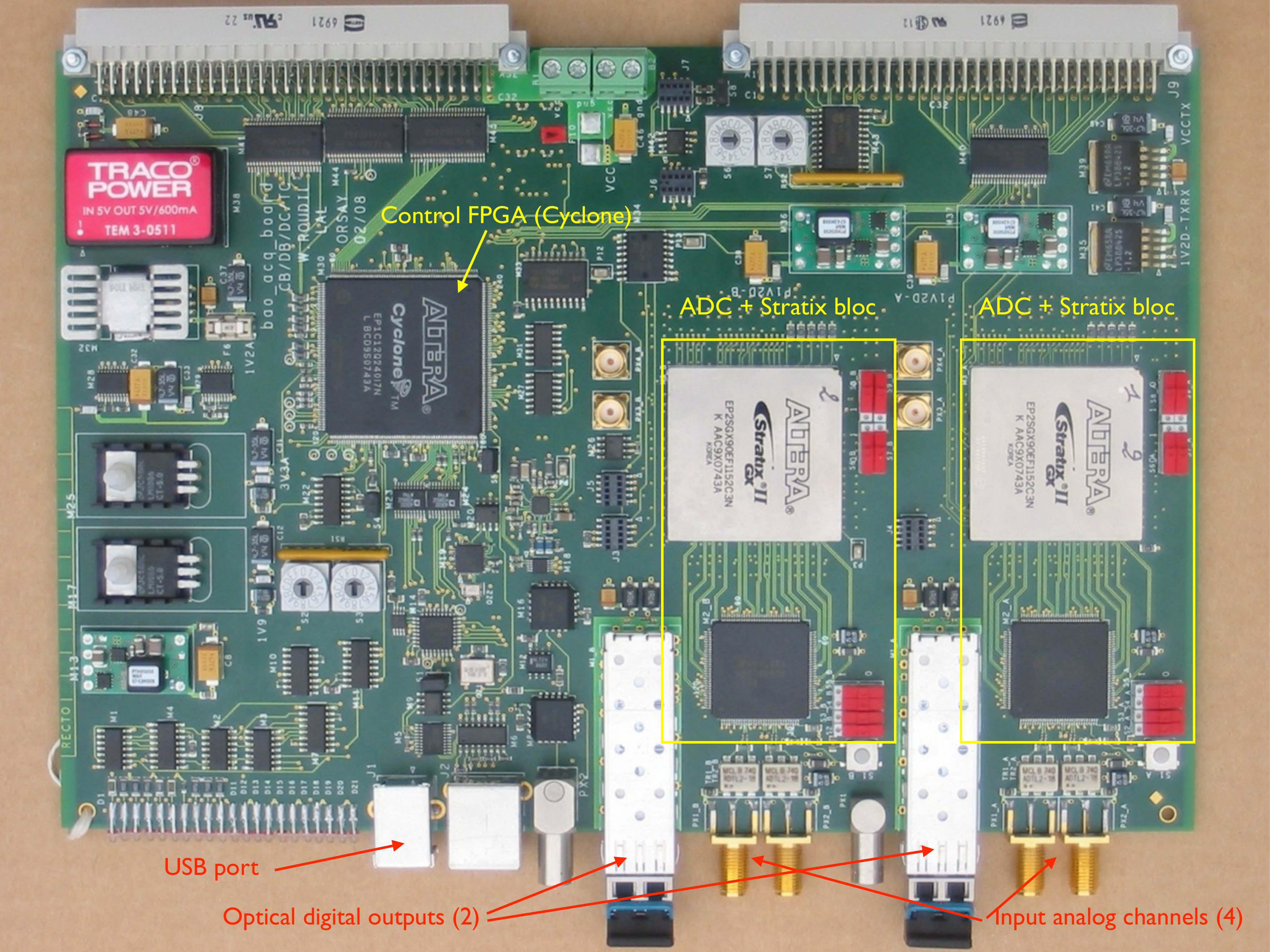}
}
\vspace*{-5mm}
\caption{Carte de numérisation (4 voies à 500 MHz) et de filtrage numérique 
des signaux développée pour le projet BAORadio. }
\label{figadcboard}
\end{figure}

\section{Conclusions, perspectives}
La cartographie \`a 3 dimensions de la brillance en \HI par la méthode de la mesure de la température de brillance
est une méthode prometteuse pour des relevés cosmologiques de la distribution de 
matière jusqu'à $z \lesssim 3$, et pour contraindre l'énergie noire en utilisant les Oscillations 
Acoustiques Baryoniques (BAO). 
Un instrument multilobes (FPA ou interféromètre)  ayant un grand champ de vue instantané
(10-100 deg$^2$) et une bande large en fréquence (quelques centaines de MHz)
 devra être utilisé afin d'obtenir une sensibilité suffisante pour le sondage d'un grand volume 
de l'univers à 21 cm.
La chaîne électronique présentée ici a été conçue pour équiper un système multi-récepteurs,
comprenant plusieurs centaines de voies de numérisation. L'électronique analogique et les cartes 
de numérisation peuvent être réparties sur une surface de plusieurs hectares, les signaux numériques
étant acheminés par fibre optique vers un système central de traitement de données au vol.
Ce système de traitement comprendra une grappe de calculateurs reliés par un réseau haut-débit,
équipés de cartes de traitement à base de GPU ou FPGA. Le coût de la chaîne électronique et du système 
central de traitement de données devrait se situer dans une fourchette de 3000 à 5000 euros par voie 
de numérisation.

\end{document}